# Development of a System Vulnerability Analysis Tool for Assessment of Complex Mission Critical Systems

Matthew Tassava, Cameron Kolodjski, Jeremy Straub
Institute for Cyber Security Education and Research
North Dakota State University
1320 Albrecht Blvd., Room 258
Fargo, ND 58108
Phone: +1-701-231-8196
Fax: +1-701-231-8255
Email: matthew.tassava@ndsu.edu, cameron.kolodjski@ndsu.edu, jeremy.straub@ndsu.edu

**Abstract**

A system vulnerability analysis technique (SVAT) for complex mission critical systems (CMCS) was developed in response to the need to be able to conduct penetration testing on large industrial systems which cannot be taken offline or risk disablement or impairment for conventional penetration testing. SVAT-CMCS facilitates the use of known vulnerability and exploit information, incremental testing of system components and data analysis techniques to identify attack pathways in CMCSs.  This data can be utilized for corrective activities or to target controlled manual follow-up testing.  This paper presents the SVAT-CMCS paradigm and describes its implementation in a software tool, which was built using the Blackboard Architecture, that can be utilized for attack pathway identification.  The performance of this tool is characterized using three example models.  In particular, it explores the path generation speed and the impact of link cap restrictions on system operations, under different levels of network size and complexity.  Accurate fact-rule processing is also tested using these models. The results show significant decreases in path generation efficiency as the link cap and network complexity increase; however, rule processing accuracy is not impacted.

**Keywords:** system vulnerability analysis, mission critical systems, industrial systems, Blackboard Architecture, penetration testing, attack pathways

## 1. Introduction

Larger systems and networks, such as most complex mission critical systems (CMCS), have numerous potential attack surfaces and points of vulnerability.  As the size of these systems grows, so does the likelihood of complex viable attack pathways in the system. System updates and other changes over time introduce the possibility of new attacks as well as the likelihood of an older vulnerability being reintroduced.  For many CMCSs, taking the system offline for active penetration testing – particularly with the frequency required to test the numerous changes that may be made to the system due to updates, patches and functional technology changes – is not feasible.  Some systems cannot risk disablement or impairment by a penetration testing process that creates the very issues that security testing seeks to prevent.

With large systems and networks, it can be difficult to locate potential vulnerabilities, whether they are newly created or have been there since day one. The ability to find all possible paths an attacker can take, depending on what and where they exploit, can help testers identify these potential problem areas. Being able to see which devices have vulnerabilities and what attackers can reach after exploiting said vulnerabilities provides valuable system security intelligence. This analysis can help establish an in-depth understanding of where security risks exist and what may need to be corrected within the system. It can also facilitate the pre-implementation analysis and security assessment of proposed changes before they are implemented in the production environment for adversaries to potentially exploit.

This paper presents the system vulnerability assessment technique (SVAT) for CMCSs and an implemented software application that is designed to automate the process of discovering points of potential vulnerabilities in a complex system. It does this by identifying and analyzing all attack pathways in the system and, thus, all possible vectors which might be utilized by a malicious actor. An abstract representation of the devices in the system is developed and can be progressively augmented with additional information as it is collected. This model is representative of the actual system and is used to perform logic-based vulnerability exploitation possibility exploration and analysis of the pathways that attackers could potentially traverse in the real-world system.

This paper continues with the presentation of prior work, in several areas, which the current work builds upon. Then, an overview of the SVAT-CMCS system and its use is presented. Following this, the results of experimentation, using three example models, are presented. Next, the results of analysis of SVAT's speed and logical operations are discussed. Finally, the paper concludes and discusses potential areas of relevant future work.

## 2. Background

This section provides an overview of prior work in several areas which serve as a foundation for the work presented herein. First, path-based cybersecurity assessment methodologies are discussed. Then, prior work on security testing automation is reviewed. Finally, the Blackboard Architecture is discussed.

### *2.1. Path Based Cybersecurity Assessment Methodologies*

Cybersecurity assessment models, according to Strom, et al. [1], can be classified into three levels: high, mid and low. Attack, defense and threat trees fall under the lowest level and can model down to individual exploits or attacks. The MITRE ATT&CK model is a mid-level framework [1] and the Lockheed Martin Cyber Kill Chain and STRIDE frameworks are both high level models [1].

Attack, defense and threat trees utilize a common tree structure for cybersecurity modeling. Attack trees, as expected, model "attack strategies against a system" [2]. As most modern attacks require exploiting multiple vulnerabilities [3], attack trees can be used to identify the requisite capabilities for attacks and they allow attackers (and those seeking to understand attacks to defend against them) to take a modular approach [3]. Problematically, attack trees don't consider potential countermeasures that defenders could deploy to attempt to prevent attacks [2].

Defense trees fill this gap [4]. They start with an attack tree and are then created through "adding a set of countermeasures to the leaves of an attack tree" [4]. Defense trees allow attackers and defenders alike to predict which attack and defense strategies could be most effective [5].

Another similar concept, attack response trees, also include both attacks and responses; however, they suffer from "state-space explosion" issues [6]. Attack countermeasure trees [6] build on the defense tree concept, allowing defenses to be located throughout the tree, as opposed to defense trees which have leaf node defenses.

Threat trees are a higher-level tree structure which are used for planning. Their goal is to "identify how and under what condition threats can be realized" [7], which can be used to estimate risks and to determine what countermeasures are needed. A threat tree is created using a four-phase modeling process [7]. It begins with system decomposition. After this, threat analysis is performed for each component and then risk analysis and prioritization of threats are conducted.

Another path-based model is the MITRE ATT&CK framework [8]. ATT&CK is threat-based and includes seven operational phases, ten tactic components and seven analytics phases [8]. The seven phases of the operations are "recon", "weaponize", "deliver", "exploit", "control", "execute" and "maintain" [8]. The control, execute and maintain phases also include a set of tactics [8] which are grouped into ten categories. These are "persistence", "privilege execution", "defense evasion", "credential access", "discovery", "lateral movement", "execution", "collection", "exfiltration", and "command and control" [8]. Each of these categories has between 8 and 29 individual tactics, which range from environment-specific attacks to broader concepts. Some tactics fall under multiple categories.

The Lockheed Martin Cyber Kill Chain was introduced in 2011 and is based on the concept of a military kill chain [9]. Its fundamental concept is that "if any stage of the kill chain is blocked then the attack will not be successful" [9]. It has served as a "mental process guide" for developing various forms of protection [9]. The Cyber Kill Chain's seven phases and very similar to those used by the MITRE ATT&CK framework for its operations. They are "reconnaissance", "weaponization", "delivery", "exploitation", "installation", "command and control", and "actions on objectives" [9]. Notably, the Cyber Kill Chain lacks the "maintain" phase that the ATT&CK model has. Cyber Kill Chain ends with the "detonation stage" during the "actions on objectives" phase [9].

The final path-based model that will be discussed is the STRIDE framework, which was developed by Microsoft [10]. This is a five-phase method that is similar to threat trees. The steps are "decomposition", "create data flow diagrams", "analyze the data flow diagrams for threats", "identification of vulnerabilities based on these threats" and "develop mitigation approaches" [10]. These steps are data-driven and notably different from the MITRE ATT&CK and Cyber Kill Chain frameworks. The STRIDE framework was designed to address six security threats: "spoofing", "tampering", "repudiation", "information disclosure", denial of services", and "elevation of privilege" [10].

### *2.2. Automation of Security Testing*

Evaluating the security of computing systems becomes more difficult as they grow larger and have more frequent changes. Despite best efforts and mitigating known vulnerabilities, systems remain potentially vulnerable to attacks of unforeseen types [11]. Some tools utilize an attack library looking for known issues, which may be overly simplistic and readily become outdated. Problematically, to fully test complex systems, the evaluation of numerous potentially complex exploits may be needed to identify potential vulnerabilities [12]. Testing also runs the risk of damaging critical systems, which are demonstrated to be insecure, during testing [13]. Mission critical system operators may fear causing the very type of outage that they are seeking to prevent during testing [14].

A variety of relevant automation tools have been developed. Examples include tools for testing cloud applications [15,16], mobile devices [17], wireless internet [18], Blockchain contracts [19], and web services [20–23]. A number of approaches to this testing have been proposed including using static and dynamic analysis [24], threat models [25], and ontologies and big data techniques [26]. Technologies such as Petri nets [27], fuzzy classifiers [28], expert systems [29], automated pathfinding [30], agent-based modeling [31], and reinforcement learning [32–36] have been utilized.

Automation tools can increase the breadth and depth of testing activities while reducing time and costs. However, Stefinko, Piskozub and Banakh [37] contend that "manual penetration tests are still more popular and useful". This is, perhaps, owing to the fact that tools lack humans' ingenuity and capability to identify new types of attacks and vulnerabilities on the fly. The development of tools that can adaptively develop new attack types is a key area of potential future work. Automation tools, even in the interim, can have a key role in facilitating test implementation and the rapid re-testing of previously tested attacks to validate that they have not been reintroduced.

### 2.3. Blackboard Architecture

The Blackboard Architecture was proposed by Hayes-Roth [38] for use in decision making. It was based on prior work on a speech recognition system [39]. Blackboard Architectures utilize the rule-fact network structure commonly found in expert systems [40], which trace their lineage back to the Dendral [41] and Mycin [42] systems in the 1960s and 1970s. The Blackboard Architecture adds actions [43], which can be used to interact with the system's operating environment. It has been used for numerous applications such as robotics [44,45], medical image interpretation [46] software testing [47] and mathematical proofs creation [48].

Prior work discussed the efficacy of using the Blackboard Architecture for cybersecurity applications. Its use for autonomous attack command [49] and implementing path-based frameworks [50] were both illustrated.

### 3. Overview

This section provides an overview of how the SVAT operates. First, the components of its operating network – the objects that define the system being tested and possible attacks – are discussed. Then, the SVAT software is briefly discussed. Finally, the operations of the SVAT system and software are demonstrated using a model network.

### 3.1. Network Components

The SVAT network is made up of facts, generic rules, common properties, containers, links, reality paths, and variants. In the software implementation, all components, except the paths and variants, need to be imported by the user in a predefined format so that the application parser can load the objects into the network. These components are built by a supporting tool.

Facts are a piece of information and hold a value of either true or false. Each fact is unique to its container. However, facts can be tied to a common property that describes what it is (e.g., *'ethernetConnected'* indicating that the device is connected to an ethernet network or *'isAdmin' indicating whether admin privilege is true or false*). While facts can be associated with only a single common property, each common property can have multiple facts associated with it. A common property, thus, is simply a description used by any number of facts. For example, many devices, which are represented by containers, have admin privilege capabilities and so the common property of *'isAdmin'* can be referenced by multiple containers' facts to indicate whether the given container's admin privilege is true or false. Figure 1 depicts the data structure for facts and common properties. Note that the fact has a true or false value, an ID that links it to a common property, if it is not null, and a description string. The common property has an ID value, which is referenced by linked facts, and a description string.

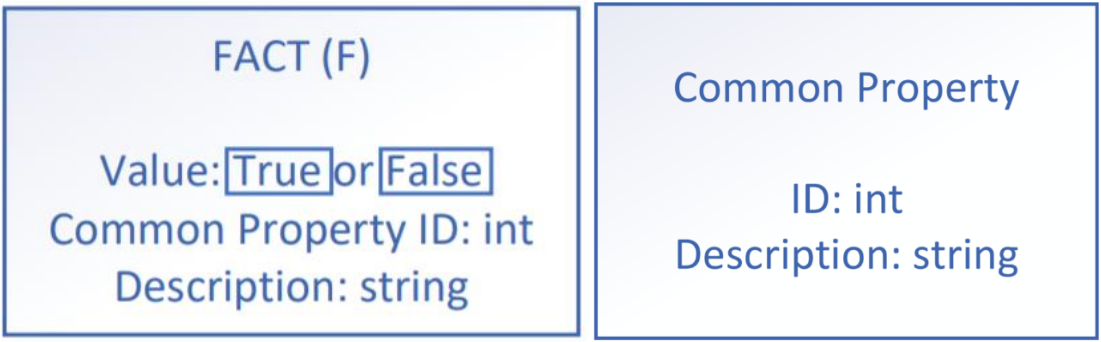

Figure 1. Depiction of a fact and common property.

Containers are the nodes in the network that are used to represent computing systems, software and other hardware such as peripheral devices, printers and routers. The containers store facts that represent information about the system, such as configuration details. The container data structure is shown in Figure 2, with a single container storing multiple facts and its description.

These facts are used as part of the virtualized security assessment testing process to match rules that represent vulnerability exploitation techniques. Rules, which are triggered by the presence of facts matching their pre-conditions, can change the accessible pathways within the network, representing the impact of an exploit being conducted.

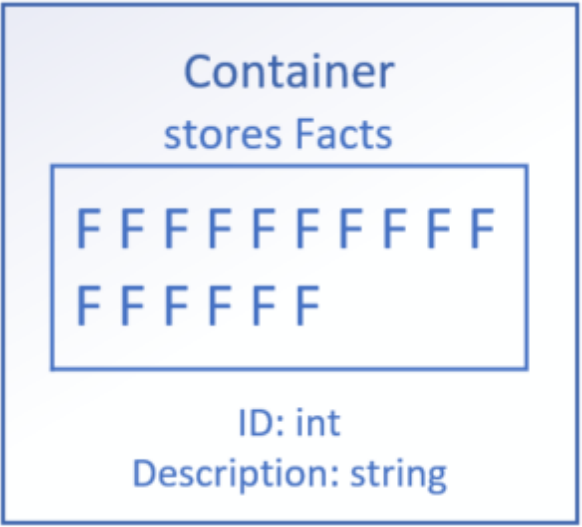

Figure 2. Container storing facts.

Links are objects that establish a one-to-one connection between two containers. A container can have multiple incoming and outgoing links. One field of the link holds the ID of the first entity it is connected to, and another holds the ID of the second entity. During traversal, if a link is found, the implementing software will find and possibly run the generic rule or rules associated with the connected containers' stored properties.

Generic rules are rules that can operate with any containers that are connected by a link and have their pre-conditions satisfied. The link data structure is depicted in Figure 3. Links contain an ID, description, the origin container's ID (Entity1ID), and their destination container's ID (Entity2ID).

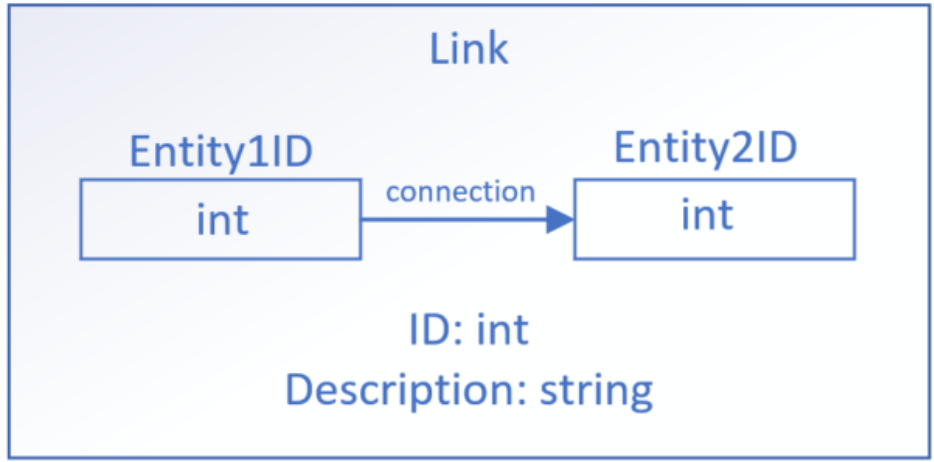

Figure 3. Link between two entities.

GenericRules are key to the operations of the technique. They are rules that operate on any pair of linked containers, if the facts in the containers on both sides of the link satisfy the requirements for the rule to be triggered. If the rule is triggered, alterations to facts stored in the containers are made as defined by the rule's post conditions. The data structure for a generic rule is shown in Figure 4. The Generic Rule has an ID, a description, and any number of pre/post conditions for the start and end containers.

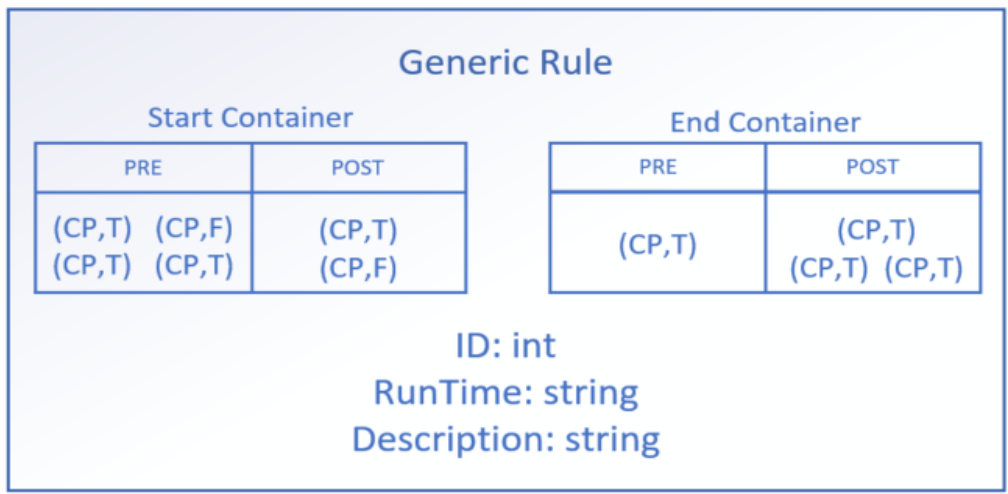

Figure 4. Depiction of a generic rule definition.

Reality paths are used to keep track of visited containers, links, and altered facts after rules have been run. When containers and facts are added to the path chain they are termed variants, in an effort to keep original containers and facts separate from those that are part of a path. This allows the software system to generate all possible pathways, even those that require facts in different states than they are currently in during the search iteration. These variant containers and variant facts are, thus, key to identifying all possible attack pathways. The data structure for a reality path is shown in Figure 5, which also shows how variants are assessed to see if generic rules' preconditions are met. As shown in Figure 5, reality paths have associated containers and facts.

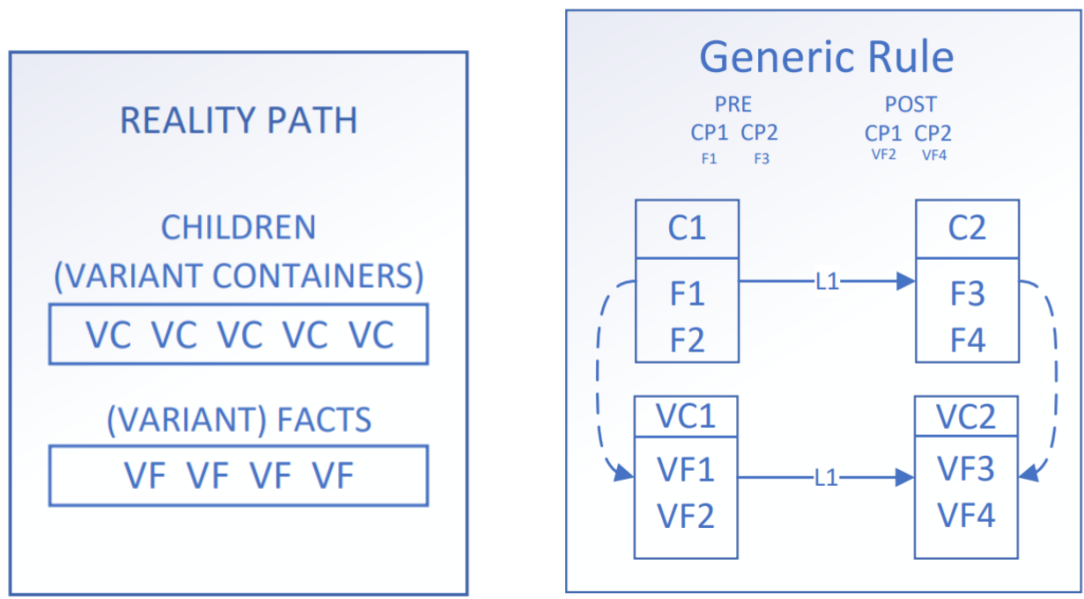

Figure 5. Depiction of a reality path.

*3.2. SVAT*

The SVAT is implemented by a software system that has three main components: the Software for Operations and Network Attack Results Review (SONARR), the Target Representation Development Module (TRDM), and the Attack Definition Module (ADM).

SONARR imports data and implements a network-traversal algorithm. It is used to ascertain the impact of attacks on the modeled system. SONARR identifies every possible attack pathway, which is not prevented by the constraints of the loaded system, that an attacker could use. It does this using the provided network of containers (which represent hardware systems), links (which represent device-to-device connectivity), the rules and facts that define system operations and the generic rules which represent attacks.

TRDM is used to create the containers, facts, properties, and links that are used by the system. The output from TRDM is imported into SONARR and ADM.

ADM is a system for creating the generic rules used by the system. These rules represent potential attacks which have defined pre- and post-conditions. Because of this, they require properties to be stored in specific formats in link connected containers, to satisfy rule pre-conditions and be modified by rule operations. These properties are loaded from TRDM.

## 3.3. Example Networks

This section uses two examples to illustrate the operations of the SONARR system. Properties are used to represent software (e.g., Windows 10, Ubuntu), configurations (e.g., administrator, firewall enabled, local network), and other information about a system (container). The example presented in Figure 6 includes a vulnerability that sets a container's compromised property to true when Win10 is true, FirewallEnabled is false, compromised is false, and either EntryPoint or CompTraversed is true. If a container lacks a firewall and is on the local network, CompTraversed is set to true to show that it is part of an infected path. CompTraversed demonstrates how common properties and rules can be used to pass information through paths to simulate traversal. Notably, this is a very simple system to facilitate easy understanding. Actual systems will typically be far more complex.

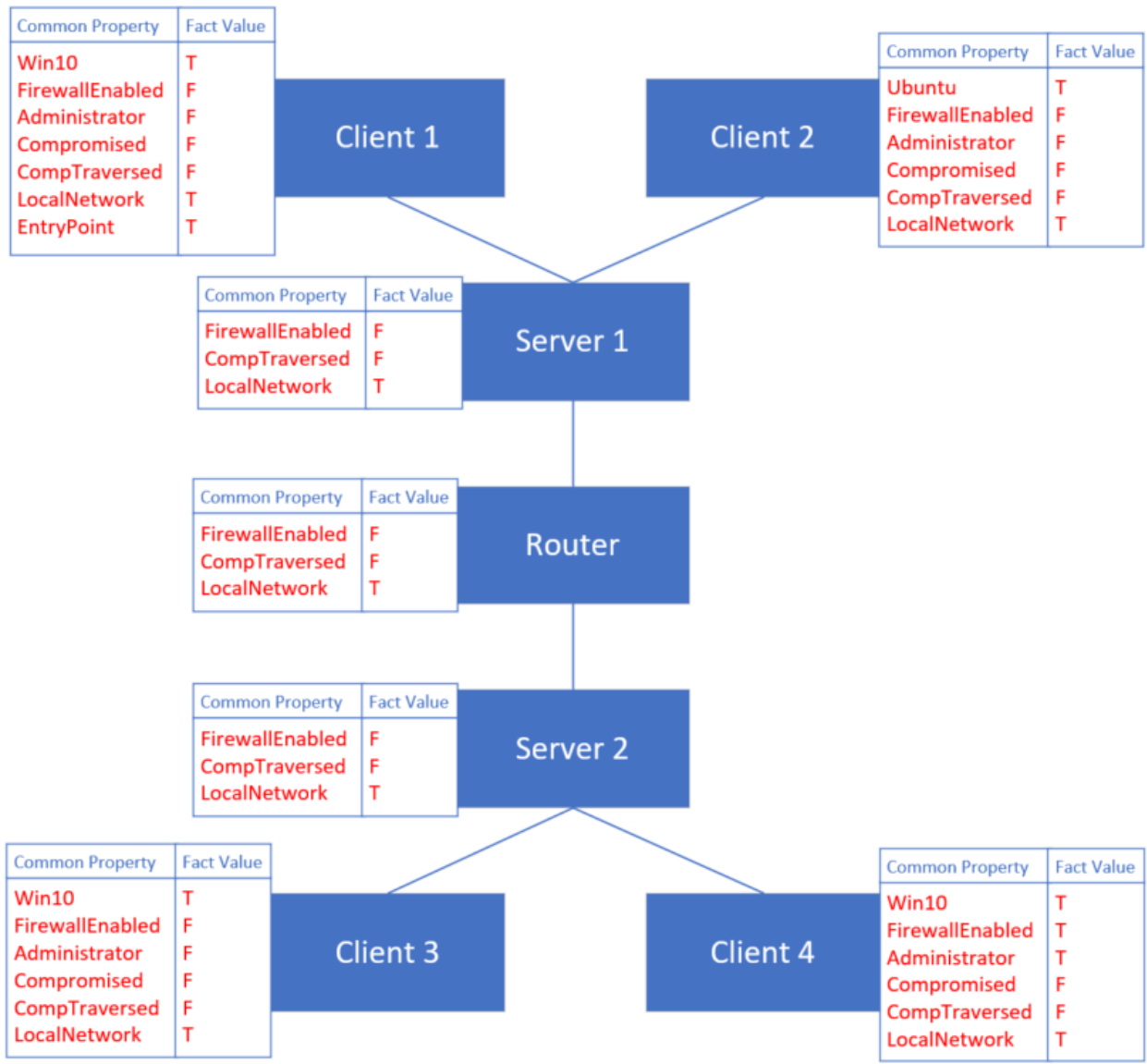

Figure 6. Example network diagram.

Figure 6 is an example of a simple network that could be loaded into SONARR. Every filled rectangle (clients, server and router) would be implemented as a container, and every connecting line would be implemented as a bidirectional link (which is implemented as two unidirectional links—one in each direction). Example facts are provided for Figure 6 and are shown in red. They consist of a common property name and a value.

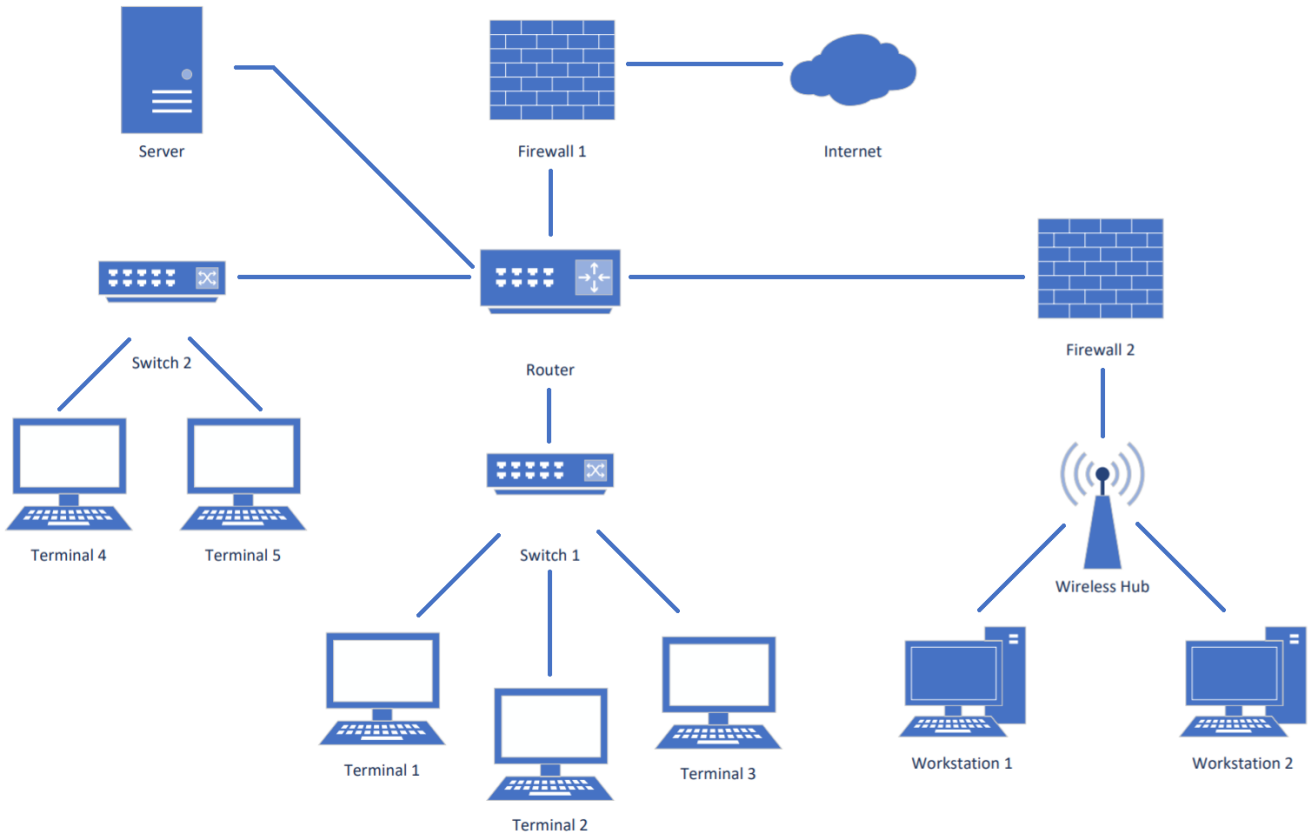

https://online.visual-paradigm.com/diagrams/templates/network-diagram/office-network-diagram-example/

Figure 7. Example office network

Figure 7 presents an example of a more complex network, like might be found in a small office environment. An implementation of this as a system network is presented in Figure 8. Like in Figure 6, each line is a bidirectional link. Systems are shown, instead of SVAT assets, in Figure 6.

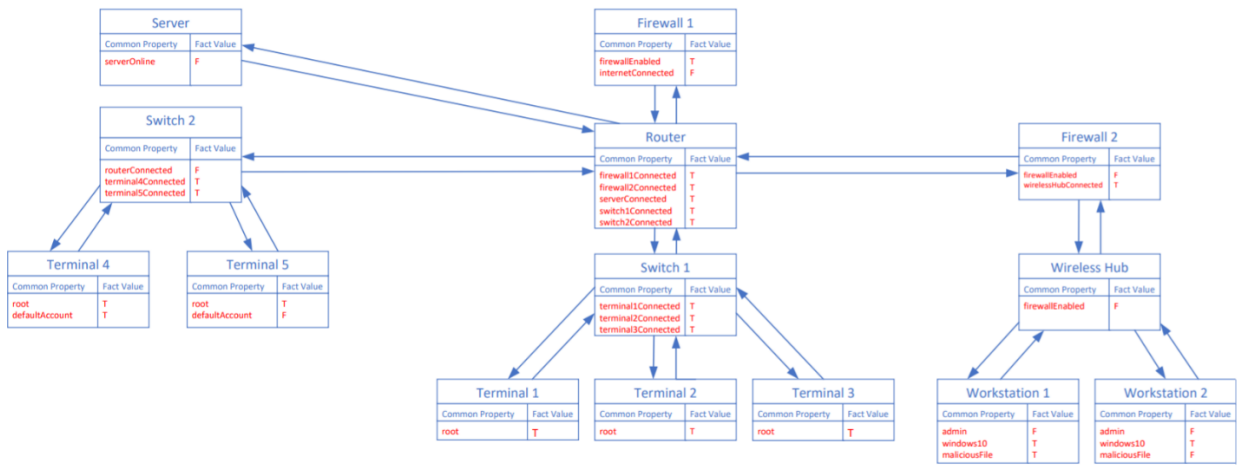

Figure 8. Example office network with facts and common properties

Each container would hold relevant facts as they pertain to the entity represented by the container and its interaction with other containers in the network. These facts would reference common properties that the user would then use to create rules to be evaluated during traversal. For example, a "Firewall" container may have a property "FirewallEnabled" that is referenced by a rule that simulates the movement of packets from the internet into the local network.

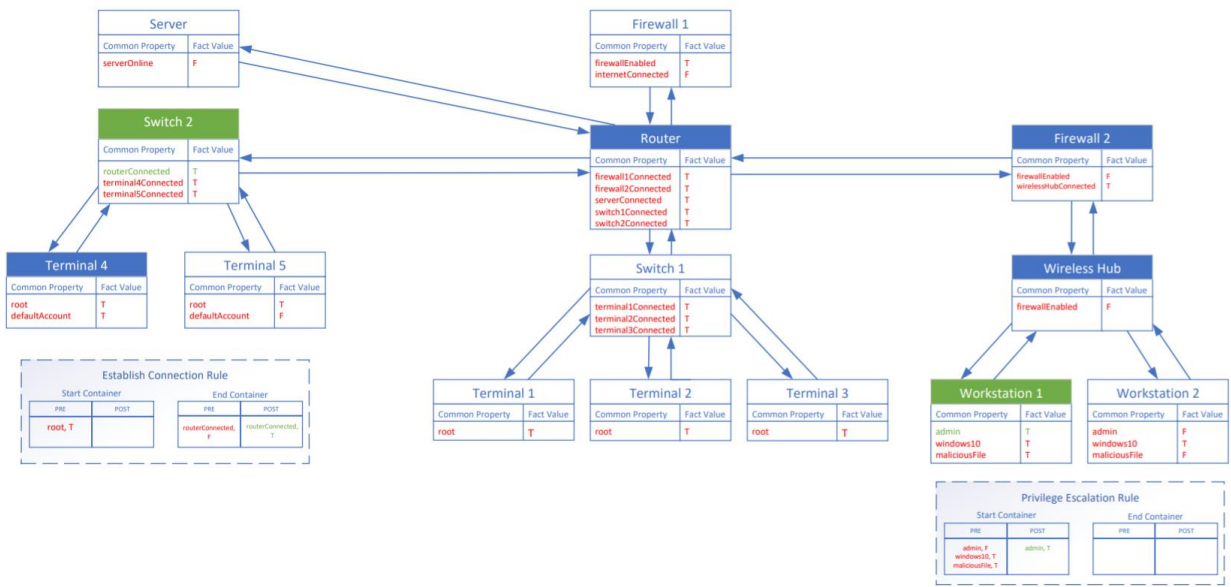

Figure 9. Example office network with altered facts after rules are triggered

Figure 8 is a simple example of how the system shown in Figure 7 could be implemented in SONARR with facts, and common properties. Figure 9 shows the state of the system shown in Figure 8 after rules are triggered simulating an attack scenario. Starting at Terminal 4, this scenario has the attacker use a default account with root privileges as an entry point to the system. They then move to Switch 2, triggering a rule representing the attacker establishing a connection to the Router. Then they move through Firewall 2, the Wireless Hub, and finally reach Workstation 1 where a malicious file is located. Reaching this stage triggers a rule representing a privilege escalation attack using the malicious file and the Windows 10 fact, changing the admin fact in Workstation 1 from false to true.

To implement and evaluate these scenarios, containers, facts, and links are created in TRDM. Containers are added by specifying a description. Links are added by selecting two containers then choosing whether the connection will be unidirectional or bidirectional. A unidirectional link is a single link from one container to the other. The bidirectional option creates two link objects that work in opposite directions.

Attacks are then defined in ADM by selecting properties and rules. Properties can be selected as preconditions from each container that must be met for the rule to run, and postconditions whose value will change if the preconditions are met and the rule runs. A description for the rule can also be added.

Once the containers, facts, links, rules, and common properties have been loaded into SONARR, a start and end container are specified and the system can be run to generate all valid attack paths (limited by the loaded constraints).

A link cap setting can also be specified to indicate whether the user wants to consider paths that require links to be traversed more than once. In some cases, links' generic rules may be needed to set or reset values, after they are changed by other generic rules, in order to produce fact configurations that satisfy generic rule pre-conditions. Thus, for some systems preventing re-traversal may exclude some valid paths. The valid paths—paths which start at the start container and end at the end container—are provided to the user.

SONARR has an additional feature of a filter mechanism that can be used to narrow down its output to focus on specific scenarios of interest. When a filter is applied, the list of paths outputted by SONARR is reduced to paths where for at least one variant of each container in the path satisfies all of the filter's constraints. Filters reference containers, and multiple filters can be applied. However, each container can only be referenced by one filter. Each filter can have as many constraints as there are common properties. For example, the model shown in Figure 6 could use the filter mechanism to specify that the common properties "Compromised" and "Administrator" must be true for Client 1, Client 2, Client 3, and Client 4. When this filter is applied, the list of paths outputted by SONARR is reduced to paths where, for containers Client 1, Client 2, Client 3, and Client 4, at least one variant in the path, has the common properties "Compromised" and "Administrator" with both true.

### 4. Experimentation

Experimentation was conducted to demonstrate system efficacy and to answer two research questions that are salient to the operations of the proposed system. First, the overall system was validated to work as has been previously described, herein. Next, the complexity and speed of network path chain generation was characterized relative to the link cap value selected. Finally, the accuracy of operations and fact value alteration results were verified manually for networks implemented with rules against expected results.

*4.1. Validation of Operations*

Before the test models were created, the individual network components were tested to verify that they functioned correctly. Containers, links, and facts were developed together and tested for functionality during development. Variables storing these components were monitored, using debug tools, and the output text for each network traversal session provided formation feedback of the path chains. Containers and their facts could be seen in both the variables and the output, showing the storage of facts within their appropriate containers. Proper pathway generation over links was also verified in the output path chains (e.g., C001,L001;C002UL002|F001T,F002F).

Generic rules and common properties started with testing on small networks. Tests ensured that pre-condition referencing was correct, appropriate rule triggering occurred at given locations in the network, and that post-condition alterations were made, through the use of debug tools. The post condition alterations were added to the path chain output facilitating verification of the rules triggered during a traversal (e.g., C001,L001,R001;C002UL002|F001T,F002T). This format of path chain output gives a detailed snapshot of each container-link-container traversal on a line-by-line basis.

Each test model in the experiments presented herein outputted these path chains. These were used to validate that the appropriate rules were triggered at proper locations, fact alteration was consistent with the rules triggered, and that the expected sequence of containers was traversed.

*4.2. Link Cap Impact*

The link cap is a limit for the traversal of each link within a path chain that prevents excessive looping. Without the link cap, traversal between two containers could loop infinitely. This is impractical, and unlikely to result in unique solutions, beyond a certain level of repetition, so the link cap was implemented. Depending on the size of the network model, increasing the link cap will increase the amount of unique path chains. It also increases the time to generate paths, thus decreasing system efficiency.

To conduct experimentation regarding the impact of the link cap, three different model networks were developed. The first model has three containers that each have two exiting links, one to each of the other containers. The second model has four containers, each of which has two exiting links connected to two of the other three containers. Finally, model 3 is the same as model 2, except that each container has an additional exit link connecting to the container that it wasn't connected to in model 2. Diagrams of each model are presented in Figures 7 to 9. Speed experiments, which don't run any logic assessments using rules (they are strictly testing the efficiency of finding links and generating unique path chains through the network) were conducted.

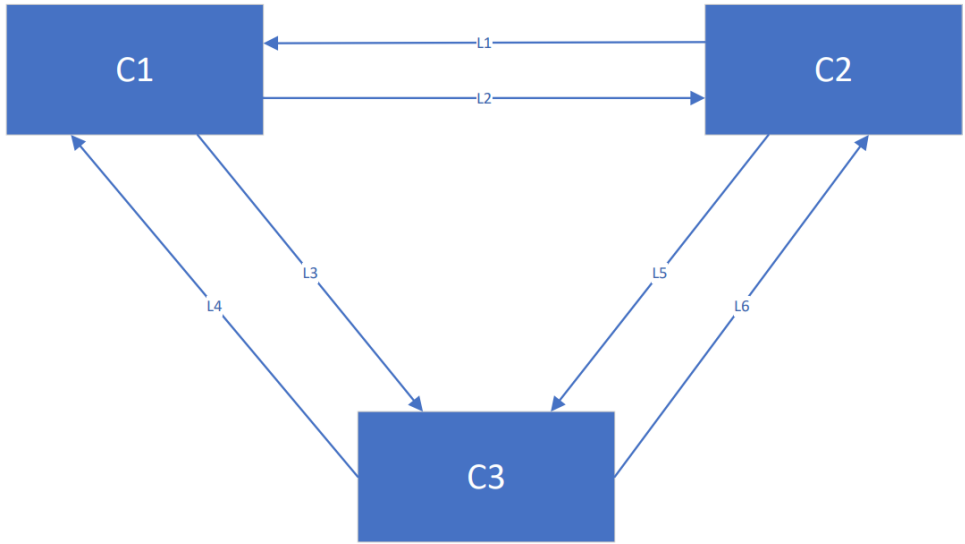

Figure 10. Model 1 diagram

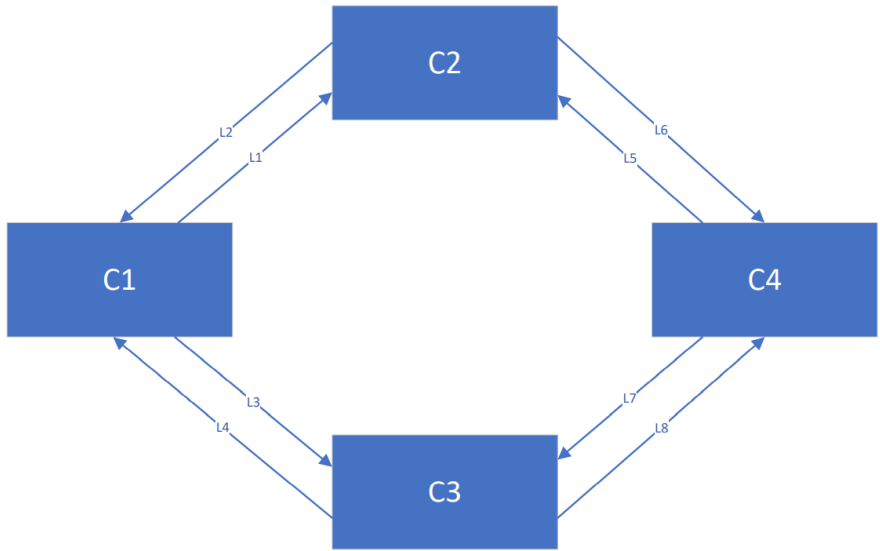

Figure 11. Model 2 diagram.

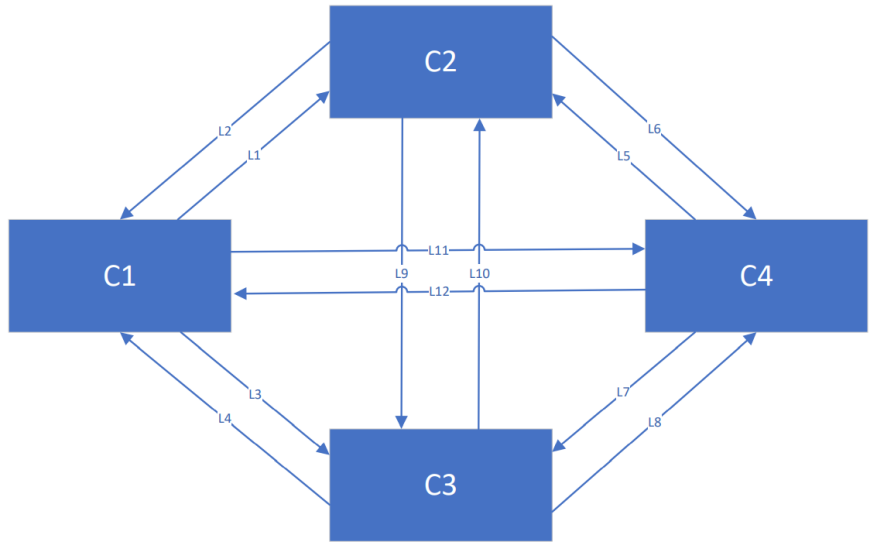

Figure 12. Model 3 diagram.

The rule-fact configuration for model 1 includes two rules that analyze properties which are in all three containers. Container 1 contains the first fact, container 2 contains the second, and container 3 contains the third. Rule 1's pre-condition requirements check the properties associated with facts 1 and 2 while rule 2 checks the pre-condition requirements against properties associated with facts 2 and 3. Table 1 presents these facts and Figure 13 depicts the network configuration and the triggered rule-caused changes that it creates.

Table 1. Model 1 Rules.

| Rule # | Start Container "Pre" Properties | End Container "Pre" Properties | Start Container "Post" Properties | End Container "Post" Properties |
| --- | --- | --- | --- | --- |
| R001 | P1:T | P2:F | --- | P2:T |
| R002 | P2:T | P3:F | --- | P3:T |

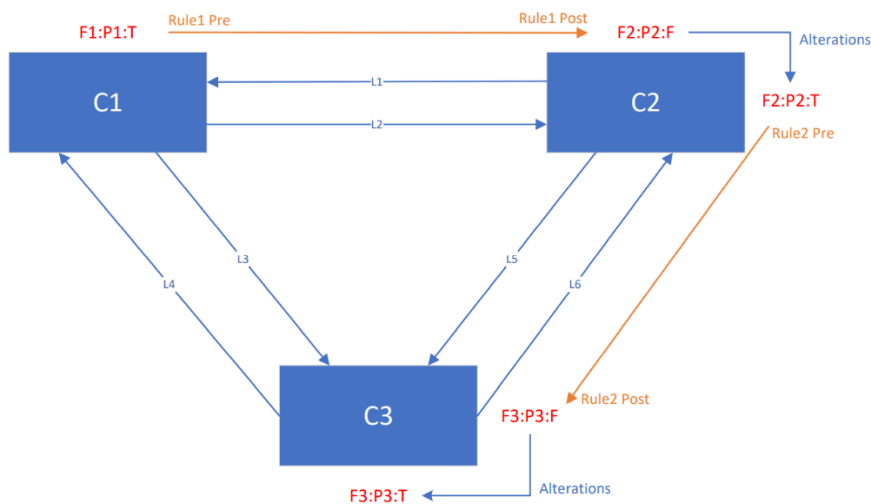

Figure 13. Model 1 rule diagram.

Similar to Table 1 and Figure 13, which contain relevant information about the first network model, Table 2 and Figure 14 present a list of the rules and visual depiction of the second network model. This network contains four containers, eight links and four rules.

Table 2. Model 2 Rules.

| Rule # | Start Container "Pre" Properties | End Container "Pre" Properties | Start Container "Post" Properties | End Container "Post" Properties |
|---|---|---|---|---|
| R001 | P1:T | P2:F | --- | P2:T |
| R002 | P2:T | P3:F | --- | P3:T |
| R003 | P3:T | P4:F | --- | P4:T |
| R004 | P4:T | P5:F | --- | P5:T |

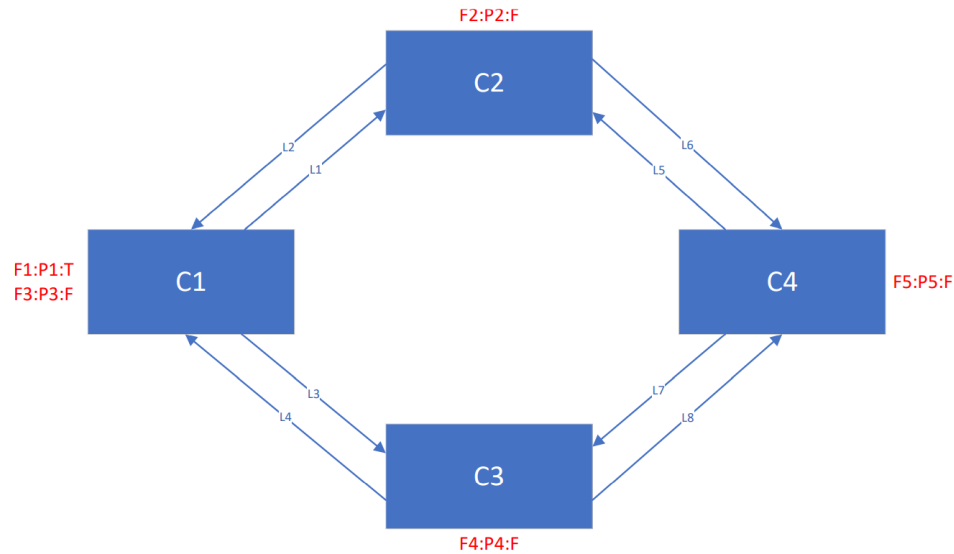

Figure 14. Model 2 rule diagram.

The third network model is presented in Table 3 and Figure 15. The third network model contains four containers, twelve links and six rules.

Table 3. Model 3 Rules

| Rule # | Start Container "Pre" Properties | End Container "Pre" Properties | Start Container "Post" Properties | End Container "Post" Properties |
|---|---|---|---|---|
| R001 | P1:T, P2:T | P5:F | --- | P5:T |
| R002 | P5:T, P6:T | P3:F | --- | P3:T |
| R003 | P3:T | P7:F | --- | P7:T |
| R004 | P7:T, P8:F | P5:T, P6:T | P8:T | --- |
| R005 | P7:T, P8:T | P4:F | --- | P4:T |
| R006 | P4:T | P9:F | --- | P9:T |

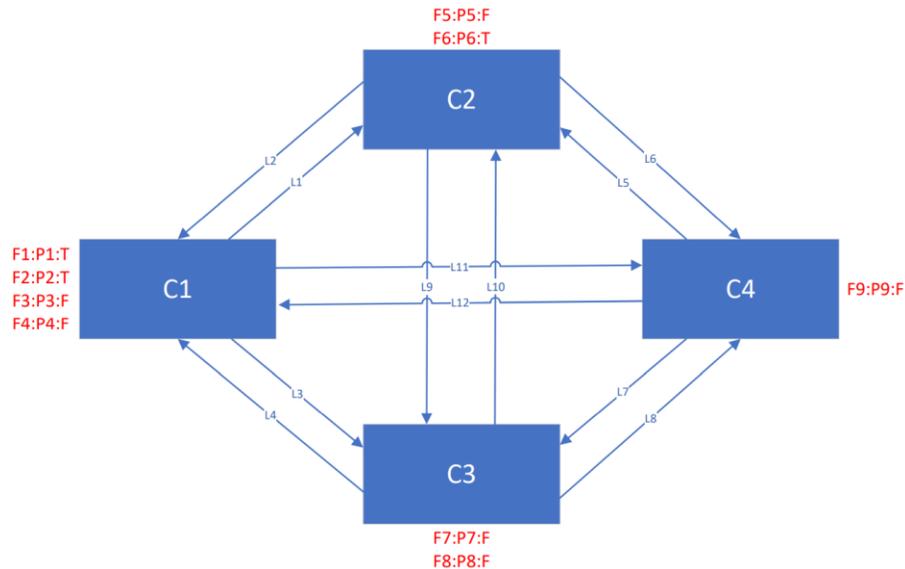

Figure 15 - Model 3 rule diagram.

**5. Data Analysis**

This section presents the results from the experimentation that was described in the previous section. For each network model, the results from the speed characterization experiments are presented in a table for each link cap value tested. The number of final paths that were found for the test with that link cap value and the time taken to find those paths are listed.

An output listing showing the results of the rule logic experimentation is also provided for each network model. This is presented in the path chain format outputted by SONARR. These path chains, an example of which is presented in Listing 1, are easily human readable. They are read from top to bottom and left to right. Each record is a single line of text.

**Listing 1.** Example output.

```
C001UL002UL003|F001T
C001,L002,R001;C002UL001UL005|F001T,F002T
C001,L003;C003UL004UL006|F001T,F003F
C001,L002,R001;C002,L001;C001UL003|F001T,F002T
C001,L002,R001;C002,L005,R002;C003UL004UL006|F001T,F002T,F003T
C001,L002,R001;C002,L001;C001,L003;C003UL004UL006|F001T,F002T,F003F
```

Each record starts with the identifier for the user-selected start container (in this case, C001) and its outgoing links (e.g., UL002UL003 for the first record). The 'U' is a delimiter signifying an exit link for a container. After a link is traversed, a new record is written that includes all of the previously traversed containers in the path chain, the exit link traversed for each container, and any rules that were run on those links (e.g., C001,L002,R001, for the first record). The containers in a path chain are separated by semicolons and commas are used as delimiters between the links and rules of the containers. The vertical bar (|) denotes separation between the path chain and the path's facts. The path's facts lists every fact from every previous container and are used to display each fact's current value to human readers while recording fact alterations for the software to use in future rules. A new line character (which is not shown in the listing) terminates each record.

*5.1. Model 1*

The first model is most straight forward in terms of complexity and performs the best, in terms of speed. There are only three containers and two links exiting each container, so with each link cap increment there are only two new unique paths generated. As shown in Table 4, the time taken to find these path sets is less than one second for each link cap value from one to five.

Table 4. Model 1 time results.

| link cap | Final Paths Found | Elapsed Time |
| --- | --- | --- |
| 1 | 3 | < 1 sec |
| 2 | 5 | < 1 sec |
| 3 | 7 | < 1 sec |
| 4 | 9 | < 1 sec |
| 5 | 11 | < 1 sec |

The rule output is as expected when the link cap value is set to one. Starting from Container 1 and traversing to Container 3 produces three final paths and all three have the expected fact alterations. The important traversal (C001 --> C002 --> C003) displays R001 having been run between C001 and C002, and R002 having been run between C002 and C003. Those rules are expected to alter facts F002 and F003 from false to true, which occurred as expected. The output from the testing with Model 1 is presented in Listing 2.

**Listing 2.** Model 1 rule example output at link cap one (C001 → C003)

```
C001UL002UL003|F001T
C001,L002,R001;C002UL001UL005|F001T,F002T
C001,L003;C003UL004UL006|F001T,F003F
C001,L002,R001;C002,L001;C001UL003|F001T,F002T
C001,L002,R001;C002,L005,R002;C003UL004UL006|F001T,F002T,F003T
C001,L002,R001;C002,L001;C001,L003;C003UL004UL006|F001T,F002T,F003F
```

*5.2. Model 2*

The second model adds a fourth container and two additional links to the network. Path generation takes longer with this additional complexity. The lowest two link cap values take less than 1 second, like with the first model. However, the amount of time taken and the number of paths identified increases significantly with higher link cap values. Link cap values from 1 to 6 require less than one minute; however, by link cap value 8, nearly 2 hours are required. Both the number of paths and time required grow at a greater than linear rate.

Table 5. Model 2 time results.

| link cap | Final Paths Found | Elapsed Time |
|---|---|---|
| 1 | 4 | < 1 sec |
| 2 | 18 | < 1 sec |
| 3 | 68 | 1 sec |
| 4 | 250 | 3 sec |
| 5 | 922 | 13 sec |
| 6 | 3430 | 57 sec |
| 7 | 12868 | 9 min |
| 8 | 48617 | ~ 119 min |

The second model's rule operations met expectations. The key path for this run (C001 -> C002 -> C001 -> C003 -> C004) reached all of the containers, and therefore triggered all of the rules. The addition of an additional container did not hinder the generic rule implementation. An additional final path was generated, as compared to the previous network model. All of the final paths were still unique, with various alterations of fact values made, due to the rules that were (or were not) triggered. A subset of the results from the operations with network model 2 are presented in Listing 3.

**Listing 3.** Model 2 example rule output at link cap one (C001 → C004)

```
C001UL001UL003|F001T,F003F
C001,L001,R001;C002UL002UL006|F001T,F003F,F002T
C001,L003;C003UL004UL008|F001T,F003F,F004F
C001,L001,R001;C002,L002,R002;C001UL003|F001T,F003T,F002T
C001,L001,R001;C002,L006;C004UL005UL007|F001T,F003F,F002T,F005F
C001,L003;C003,L004;C001UL001|F001T,F003F,F004F
C001,L003;C003,L008;C004UL005UL007|F001T,F003F,F004F,F005F
C001,L001,R001;C002,L002,R002;C001,L003,R003;C003UL004UL008|F001T,F003T,F002T,F004T
(5 more lines not shown)
```

*5.3. Model 3*

The third model is based on the second model; however, two additional bi-directional links are added between the non-adjacent containers, as shown in Figure 12. These additions increase the total number of links to twelve. Each container is now directly connected to every other container. This addition significantly increases the completion time for all but the lowest link cap value (of 1). Only link cap values of 1 to 3 were successful in completing their runs within two hours. With a link cap value of 4, the system was not able to complete processing within two hours, at which point the testing was terminated. Because of this, higher link cap values were not tested. The time results and the number of paths identified for each link cap value, for model 4, are presented in Table 6.

Table 6. Model 3 time results.

| link cap | Final Paths Found | Elapsed Time |
|---|---|---|
| 1 | 33 | < 1 sec |
| 2 | 1027 | 8 sec |
| 3 | 39553 | 23 min |
| 4 | Did not complete | > 2 hrs |

The rule processing for the third network model also completed successfully with the results expected. This model had multiple rules (e.g., R004 and R006) which went back to previous containers after altering those containers' facts. Those newly set facts allowed future rules to be triggered, continuing the expected path chain and demonstrating the efficacy of SONARR for complex network systems. SONARR successfully produced the following expected path with each container having the desired facts after rule running:

C001 -> C002 -> C001 -> C003 -> C002 -> C003 -> C001 -> C004

The formatted path chain is too large to include, so no listing is provided for this test.

*5.4. Analysis*

The previous three subsections have demonstrated the efficacy of the SONARR system for several basic network model configurations, different levels of looping prevention enabled and different rule sets. Notably, the third network model required looping for some paths, as some facts needed to be set to alternate values for triggering required rules.

In terms of speed, as is expected, the larger the network and link cap value, the slower the program performed. A key operational cost is the requirement to track changes along with the path, as it grows in length and complexity. Irrespective of the complexity level, the path generation mechanism was shown to operate suitably and as expected. It also demonstrated being able to support paths that require looping in the third model.

As the link cap value is increased, more paths – which are based on user-selected start and end containers coupled with how many times links can be traversed – are created. Each path, despite the re-traversal (or looping) over links is unique – the path chains indicating which containers are reached and which links are used to reach them never repeat. With higher link cap values, paths become progressively more complex due to the allowed reuse of previously traversed links. Notably, this is not useful in all cases. Only the third model made use of it as part of a completion requirement. However, more complex paths were also generated for the other two networks and numerous unnecessary complete paths were also generated for the third model. The development of a more selective mechanism for determining when, and to what extent, looping is allowed is, thus, a potential area for future work to improve the performance of the system.

These experiments also demonstrated the efficacy of fact alteration for networks implemented with rules. Rules were run between applicable containers based on the common properties within each. The running of rules in selected tests allowed the alteration of facts in containers which were previously unreachable without the execution of prior rules. This capability is integral to being able to fully capture complex network systems.

**6. Conclusions and Future Work**

This paper has presented a paradigm for system vulnerability analysis for systems that cannot be taken offline for testing and risking temporary impairment is similarly unacceptable. It has also presented a supporting software tool for SVAT and analyzed the performance of the paradigm and this tool.

The SVAT tool models systems under test using a Blackboard Architecture network. It implements computing systems and software as container objects and interconnects them with network segments modeled as links. The system implements attacks as rules, which are triggered during the simulated penetration testing, as applicable, and dynamically change the environment that the SONARR software is operating within. SONARR traverses this network, performing the simulated attack and generating a comprehensive list of all possible attack paths, which are not blocked by the loaded system configuration constraints.

The utility of the proposed paradigm is directly affected by how effective the SVAT software is at implementing it. The SVAT system and SONARR software were, thus, tested to assess how they would perform under various operating conditions and network model designs. The system was shown to be able to effectively model the paradigm and its operations were demonstrated to be logically sound. Factors such as network complexity, the traversal caps on links, and rules that require looping movement, where containers are revisited, have a direct influence over the speed of path generation.

This paper also explained how networks are developed, using TRDM, and how known attacks can be modeled using ADM. Once networks are loaded and the logic that establishes unique pathways between start and end points in the system has been produced, the SONARR system can be used to simulate a penetration test on the system. With accurate network importation and attack rule creation, SONARR can be used to find weaknesses within target networks, without requiring operational systems to be taken offline.

The experimentation presented herein has shown that the addition of containers and link to models slows the system down notably. However, perhaps more impactful is the link cap setting value. Higher link cap values allow more complex looping networks to be used as part of simulated attacks, but also have a notable performance impact.

The largest and most complex model used in this study, model 3, had only four containers and 12 links between them. Despite this small size, setting the link cap to four resulted in over two hours of waiting for a completed output (which was not produced within the slightly over two-hour test window).

With further development, the SVAT system is expected to be able to handle greater link cap values and larger networks with significantly reduced runtimes. Better decision-making for when to allow (or not allow) the revisiting of network nodes will be critical to this. Other areas of future work are the development of an enhanced user interface and revisiting the network component designs to further enhance overall efficacy. These improvements could provide system users with more detailed traversal results, better control over network pre-traversal fact/rule configuration, and well-defined components that minimize redundancy while running rules during traversal.

**Acknowledgements**

This work has been funded by the U.S. Missile Defense Agency (contract # HQ0860-22-C-6003).